\RequirePackage{fix-cm} 
\documentclass[a4paper, twoside, reqno, 12pt, dvips]{amsart}

\usepackage{fixltx2e}     

\usepackage{amssymb}
\usepackage{amsmath}
\usepackage{mathrsfs, dsfont}

\usepackage{a4wide}

\usepackage{indentfirst}
\usepackage{graphicx}
\usepackage{psfrag}

\usepackage[english]{babel}
\usepackage[latin1]{inputenc}

\usepackage{color}
\definecolor{oneblue}{rgb}{0,0.0,0.75}
\usepackage[colorlinks,
            urlcolor=oneblue,
            linkcolor=oneblue,
            citecolor=oneblue,
            bookmarksopen=false,
            pagebackref]{hyperref}

\numberwithin{equation}{section}

\newcommand{\w}{\ensuremath{\mathbf{w}}}
\newcommand{\A}{\ensuremath{\mathcal{A}}}
\newcommand{\M}{\ensuremath{\mathcal{M}}}
\newcommand{\E}{\ensuremath{\mathcal{E}}}
\newcommand{\G}{\ensuremath{\mathcal{G}}}
\newcommand{\N}{\ensuremath{\mathcal{N}}}
\newcommand{\F}{\ensuremath{\mathcal{F}}}
\renewcommand{\L}{\ensuremath{\mathcal{L}}}
\renewcommand{\u}{\ensuremath{\mathbf{u}}}
\newcommand{\eps}{\ensuremath{\varepsilon}}
\renewcommand{\H}{\ensuremath{\mathcal{H}}}
\newcommand{\Hilb}{\ensuremath{\mathbf{H}}}
\newcommand{\dxi}{\ensuremath{\partial_{\xi}}}
\newcommand{\dtau}{\ensuremath{\partial_{\tau}}}
\newcommand{\od}[2]{\ensuremath{\frac{d #1}{d #2}}}

\newcommand{\oD}[2]{\ensuremath{\frac{\delta #1}{\delta #2}}}
\newcommand{\pd}[2]{\ensuremath{\frac{\partial #1}{\partial #2}}}

\begin{document}

\title[Spatial Hamiltonian Dysthe equation]{Hamiltonian description and traveling waves of the spatial Dysthe equations}

\author[F. Fedele]{Francesco Fedele$^*$}
\address{School of Civil and Environmental Engineering, Georgia Institute of
Technology, Atlanta, USA}
\email{fedele@gatech.edu}
\urladdr{http://savannah.gatech.edu/people/ffedele/Research/}
\thanks{$^*$ Corresponding author}

\author[D. Dutykh]{Denys Dutykh}
\address{LAMA, UMR 5127 CNRS, Universit\'e de Savoie, Campus Scientifique, 73376 Le Bourget-du-Lac Cedex, France}
\email{Denys.Dutykh@univ-savoie.fr}
\urladdr{http://www.lama.univ-savoie.fr/~dutykh/}

\begin{abstract}
The spatial version of the fourth-order Dysthe equations describe the evolution of weakly nonlinear narrowband wave trains in deep waters. For  unidirectional waves, the hidden Hamiltonian structure and new invariants are unveiled by means of a gauge transformation to a new canonical form of the evolution equations. A highly accurate Fourier-type spectral scheme is developed to solve for the equations and validate the new conservation laws, which are satisfied up to machine precision. Further, traveling waves are numerically investigated using the Petviashvili method. It is found that their collision appears inelastic, suggesting the non-integrability of the Dysthe equations.
\end{abstract}


\maketitle

\tableofcontents

\section{Introduction}

Mathematical models used in physics and mechanics do not always possess a canonical Hamiltonian structure. Typically, the dynamics is governed by partial differential equations expressed in terms of physically-based variables, which are not usually canonical. A transformation to new variables is needed in order to unveil the desired structure explicitly (see, for example, \cite{Seliger1968}). This is the case for the equations of motion for an ideal fluid: in the Eulerian description, they cannot be recast in a canonical form, whereas in a Lagrangian frame the Hamiltonian structure is revealed by Clebsch potentials (see, for example, \cite{Seliger1968}, \cite{Morrison1998}). Moreover, multiple-scale perturbations of differential equations expressed in terms of non-canonical variables typically lead to approximate equations that do not maintain the fundamental conserved quantities, as the hydrostatic primitive equations on the sphere \cite{Lorenz1960}, where energy and angular momentum conservation are lost under the hydrostatic approximation.

Clearly, if canonical variables can be identified, then the associated Hamiltonian structure provides a natural framework for making consistent approximations that preserve the fundamental dynamical properties of the original system, notably its conservation laws \cite{Salmon1983, Shepherd1990}. For example, consider the equations that describe the irrotational flow of an ideal incompressible fluid of infinite depth with a free surface. Their Hamiltonian description was discovered by \cite{Zakharov1968} in terms of the free-surface elevation $\eta(x,t)$ and the velocity potential $\varphi(x,t) = \phi(x,z=\eta(x,t),t)$ evaluated at the free surface of the fluid. Variables $\eta(x,t)$ and $\varphi(x,t)$ are conjugated canonical variables with respect to the Hamiltonian $\H$ given by the total wave energy. By means of a third order expansion of $\H$ in the wave steepness, \cite{Zakharov1999} derived an integro-differential equation in terms of canonical conjugate Fourier amplitudes, which is Hamiltonian and has no restrictions on the spectral bandwidth. To derive the Zakharov equation, fast non-resonant interactions are eliminated via a canonical transformation that preserves the Hamiltonian structure \cite{Krasitskii1994, Zakharov1999}. Alternatively, envelope equations for the free-surface and wave potential can be obtained via multiple-scale pertubation techniques that remove such terms \cite{Zakharov1968, Stiassnie1984}. However, the Hamiltonian structure is lost since the transformation from the Zakharov action variables to envelope amplitudes is not canonical. Nevertheless, the Hamiltonian and non-Hamiltonian versions of the Zakharov's equation are completely equivalent from a practical view-point.

The modified Nonlinear Schr\"{o}dinger (NLS) equations derived by Dysthe \cite{Dysthe1979} are also non-Hamiltonian. Using the method of multiple scales, he extended the deep-water cubic NLS equation for the time evolution of the unidirectional narrowband envelope $A$ of the velocity potential with carrier wave $\exp (ik_{0}x - i\omega_{0}t)$ and that of the potential $\phi$ of the wave-induced mean flow, to fourth order in steepness and bandwidth. Introducing dimensionless units, $t^{\prime} = \omega_{0}t$, $x^{\prime} = k_{0}x$, $A^{\prime} = k_{0}A$, and dropping the primes, $\phi$ can be easily found by means of the Fourier-transform and a single equation for $A$ can be derived as well:
\begin{equation}\label{eq:envB}
A_{t}+\frac{1}{2}A_{x}+\frac{1}{8}iA_{xx}-\frac{1}{16}A_{xxx}+\frac{i}{2}%
\left\vert A\right\vert^{2}A+\frac{3}{2}\left\vert A\right\vert A_{x}+\beta
A^{2}A_{x}^{\ast} + \frac{1}{2}iA\Hilb(\left\vert A\right\vert^{2})_{x}=0,
\end{equation}%
where\ $\beta =1/4$, the subscripts $A_{t}=\partial _{t}A$ and $A_{x}=\partial _{x}A$ denote partial derivatives with respect to $x$ and $t$ respectively, $\Hilb(f)$ is the Hilbert Transform of a function $f(x)$, and $A^{\ast}$ denotes complex conjugation (see also \cite{Janssen1983}). The form of the equation for the envelope $B$ of the free surface is similar to (\ref{eq:envB}), but the term $\beta A^{2}A_{x}^{\ast}$ becomes $-\beta B^{2}B_{x}^{\ast}$ \cite{Hogan1985, Hogan1986}. As noted by \cite{Hogan1985} such terms do not play any role in the linear stability of unifrom wave trains within the order of truncation of the perturbation analysis leading to the Dysthe equation. However, in three dimensions the $\beta$-term yields a better confinement of the instability region for the Stokes wave and the associated Dysthe equations are less susceptible to numerical instability and energy leakage \cite{Trulsen1997}. Moreover, \cite{Hogan1986} noted that the two equations for $A$ and $B$ are so similar '\textit{because of some deeper structure in the problem which is as yet unresolved}'. This may well be a Hamiltonian structure that the temporal Dysthe equations lack because (\ref{eq:envB}) is expressed in terms of envelope variables, which are not canonical. Recently, the associated canonical form (which do not contain the $\beta $ term)\ has been derived by \cite{Gramstad2011} starting from the Hamiltonian Zakharov equation with the Krasitskii kernel \cite{Krasitskii1994}. We also point out that \cite{Zakharov2010}, starting from a conformal-mapping formulation of the Euler equations derived another version of the temporal Dysthe equation, which is similar to (\ref{eq:envB}) but also non-Hamiltonian.

On the other hand, to model wave propagation in wave basins a change to a coordinate system moving at the group velocity can be used by introducing the dimensionless variables%
\[
  A=\eps v,\qquad \tau =\eps (2x-t),\qquad \xi =\eps ^{2}x,
\]
with $\eps =k_{0}a$ being the wave steepness of the carrier wave and $a$ the associated amplitude \cite{Lo1985}. As such, the temporal Dysthe equation (\ref{eq:envB}) transforms, up to the fourth order in $\eps $, to
\begin{equation}\label{eq:SK}
v_{\xi }+iv_{\tau \tau }+i\left\vert v\right\vert ^{2}v+8\eps \left\vert
v\right\vert^2 v_{\tau } + 2\eps iv\Hilb(\left\vert v\right\vert ^{2})_{\tau } = 0.
\end{equation}%
hereafter referred to as the spatial Dysthe of the envelope $v$ of the wave potential. On the other hand, the associated envelope $B=\eps u$ of the free surface satisfies (see also \cite{Trulsen1997})
\begin{equation}\label{eq:PO}
u_{\xi }+iu_{\tau \tau }+i\left\vert u\right\vert ^{2}u+8\eps \left\vert
u\right\vert^2 u_{\tau }+2\eps u^{2}u_{\tau }^{\ast } + 
2\eps iu\Hilb(\left\vert u\right\vert ^{2})_{\tau } = 0,
\end{equation}%
Both (\ref{eq:SK})\ and (\ref{eq:PO})\ can also be derived directly from the Zakharov equation \cite{Kit2002}.

In this paper, we will unveil the hidden canonical structure of (\ref{eq:SK}) and that of (\ref{eq:PO}). Following our previous letter \cite{Fedele2011}, we first review the properties of Hamiltonian systems and then introduce a gauge transformation that yields a canonical form for the spatial Dysthe (\ref{eq:PO}), and new invariants for it. As a corollary, we will also show that equation (\ref{eq:SK}) for the wave potential envelope $v$ is already Hamiltonian, as expected since (\ref{eq:SK}) is a special case of the more general result obtained for two horizontal dimensions by \cite{Carter2001}. Further, a highly accurate pseudo spectral scheme is exploited to validate the new conservation laws up to machine precision. Then, ground states and traveling waves are numerically investigated using the Petviashvili method (\cite{Petviashvili1976}; see also \cite{Lakoba2007}, and \cite{Yang2010}). Their collision is numerically investigated in order to provide new insights on the integrability of the spatial Dysthe equations.

\section{Hamiltonian description}

For the spatial evolution of canonical systems, the Hamilton's equations take the form (see, for example, \cite{Arnol'd1989, Morrison1998})%
\begin{equation}\label{eq:a}
u_{\xi }^{i}=J_{c}^{ij}\frac{\delta \H}{\delta u_{j}}, 
\end{equation}%
where $u^{i}(\xi ,\tau )$ is the $i^{th}$ variable of the dynamical vector $\u$, $u_{\xi }^{i}=\partial _{\xi }u^{i}$ denotes space derivative, $\H(\mathbf{u},\xi )$ is the Hamiltonian, and $J_{c}^{ij}$ is a skew-symmetric (symplectic) matrix. The Einstein's notation for repeated index summation is adopted, and $\delta $ denotes variational differentation. If $\H$ does not depend explicitly on $\xi $, the flow generated in the phase-space of $\u$ conserves $\H$. This follows from (\ref{eq:a})
\[
\od{\H}{\xi} = \dxi\H + \oD{\H}{u_{j}}\dxi u^{j} = \oD{\H}{u_{j}} J_{c}^{ij}\oD{\H}{u_{j}} = 0,
\]
since $\dxi\H = 0$ and $J_{c}^{ij} = -J_{c}^{ji}$. If one considers a general transformation 
\begin{equation}\label{eq:tr}
 w^{i} = w^{i}(\u),
\end{equation}
the Hamilton equations (\ref{eq:a}) in the new variables become
\begin{equation}\label{eq:c}
 \dxi w^{i} = J^{im}(\w)\oD{\H}{w_{m}},
\end{equation}
where
\begin{equation}\label{eq:er}
 J^{ij}(\w) = \pd{w^{i}}{u^{n}} J_{c}^{nm} \pd{w^{j}}{u^{m}}
\end{equation}
is the cosymplectic form (see, for example, \cite{Morrison1998}). The transformation (\ref{eq:tr}) is canonical if 
\begin{equation}\label{eq:cc}
 J_{c}^{ij}=\pd{w^{i}}{u^{n}} J_{c}^{nm} \pd{w^{j}}{u^{m}},
\end{equation}
then the Hamilton's equations (\ref{eq:a})\ are invariant, otherwise (\ref{eq:c}) is in a noncanonical form and $J^{ij}$ may be singular. However, if it satisfies certain algebraic properties, i.e. skew symmetry and the Jacobi identity, then a generalization of the Darboux theorem garanties the existence of a transformation from $J^{ij}$ to $J_{c}^{ij}$ (see, for example, \cite{Morrison1998}) for a more comprehensive discussion on canonical and noncanonical forms). Obviously, by switching $\xi $ with $\tau$ the above description also holds for the time evolution of Hamiltonian systems.

As an example, consider the finite dimensional system of an harmonic oscillator in the classical canonical variables $q(t)$ (coordinate) and $p(t)$ (momentum). This admits the canonical form 
\begin{equation}\label{eq:ho}
  \dot{q} = \pd{\H}{p} = p, \qquad \dot{p} = -\pd{\H}{q} = -q,
\end{equation}
where $\dot{p}$ denotes time derivative, and $\H = (q^{2}+p^{2})/2$. The flow in the phase-space is 'incompressible' since the divergence vanishes:
\[
\pd{\dot{q}}{q} + \pd{\dot{p}}{p} = 0.
\]
The transformation $z=q+ip$ is canonical and (\ref{eq:ho}) transforms to
\begin{equation}\label{eq:h1}
  \dot{z} = -i\pd{\H}{z^{\ast}} = -iz, \qquad 
  \dot{z}^{_{\ast }} = i\pd{\H}{z}=iz^{\ast},
\end{equation}
where $\H = \left\vert z\right\vert^{2}/2$, and $z^{\ast }$ is the complex conjugate of $z$. It is straightforward to prove that the gauge transformation 
\[
  z = w e^{i\alpha\left\vert w\right\vert ^{2}},
\]
with $\alpha$ as a free parameter, is also canonical, and (\ref{eq:h1}) remains unchanged in the new variables $w$ and $w^\ast$. On the other hand, if one considers the coordinate change
\begin{equation}\label{eq:zn}
  Q = \frac{q}{\sqrt{1 + \alpha q^{2}}}, \qquad P = p,
\end{equation}%
then (\ref{eq:ho})\ transforms to the noncanonical form
\begin{equation}\label{eq:w}
  \dot{P} = -Q\left( \sqrt{1 + \alpha P^{2}}-2\alpha P^{2}\right) ,\qquad 
  \dot{Q} = \frac{P}{\sqrt{1 + \alpha P^{2}}}.
\end{equation}
This flow does not preserve volume as (\ref{eq:ho}) does, nonetheless equations (\ref{eq:w}) are those of a disguised harmonic oscillator obtained via the noncanonical change of variables (\ref{eq:zn}) (see also \cite{Morrison1998}). The Dysthe equation (\ref{eq:PO}) shares the same roots as (\ref{eq:w}). They both come from a noncanonical transformation of a Hamiltonian system. In the following, canonical variables for the spatial Dysthe are unveiled.

\section{Canonical form of the spatial Dysthe equations}

Hereafter, we will consider the generic nonlinear equation
\begin{equation}\label{eq:we}
  u_{\xi } = -ia u_{\tau\tau} - ih\left\vert u\right\vert ^{2}u 
  - c\eps\left\vert u\right\vert ^{2}u_{\tau }-\eps eu^{2}u_{\tau }^{\ast
}-fi\eps u \Hilb(\left\vert u\right\vert ^{2})_{\tau },
\end{equation}%
with $(a,h,c,e,f)$ as a quintuplet of arbitrary real coefficients. In particular, the spatial Dysthe for the wave and potential envelopes follow from (\ref{eq:we}) with parameters $\left(1,1,8,2,2\right) $ and $\left(1,1,8,0,2\right)$ respectively. The wave action 
\begin{equation}\label{eq:Ac}
  \A = \int \left\vert u\right\vert ^{2}d\tau
\end{equation}
is conserved by (\ref{eq:we})\ since
\begin{equation}\label{eq:7}
  \dxi\left\vert u\right\vert^{2} + \partial_{\tau }R = 0,
\end{equation}
where
\[  
  R = ia(u_{\tau }u^{\ast } - u_{\tau }^{\ast }u) 
  + \frac{c+e}{2}\eps\left\vert u\right\vert ^{4}.
\]
Up to date, no other conservation laws are known for $u$. Drawing from \cite{Colin2006} (see also \cite{Wyller1998}), the invariance of $\A$ suggests the following variable change via the gauge transformation%
\begin{equation}\label{eq:9}
 w = \G(u) = u\exp (ik\psi ),
\end{equation}
where $k$ is a free parameter, and the 'stream function' $\psi $ is defined as 
\[ \dtau\psi =\psi_{\tau }=\left\vert u\right\vert ^{2}. \]

Note that $\left\vert u\right\vert^{2}=\left\vert w\right\vert^{2}$ and the wave action $\A$ is preserved in the transformation. Further, from (\ref{eq:7}) the spatial evolution of $\psi_{\xi}$ is governed by 
\[ \partial_{\xi\tau }\psi = -\dtau R, \]
so that $\dxi\psi = \psi_{\xi } = -R$. The spatial evolution equation for $w$ follows from (\ref{eq:9}) as%
\begin{equation}\label{eq:g}
  w_{\xi } = u_{\xi }\exp (ik\psi ) + ikw\psi _{\xi }=u_{\xi }\exp (ik\psi )-ikwR.
\end{equation}
From (\ref{eq:we}), $u_{\xi}$ can be given in terms of time derivatives of $u$, which by means of (\ref{eq:9}) can be expressed in terms of $w$ and $\psi $ as
\begin{eqnarray*}
  u_{\tau } &=&\exp (-ik\psi )\left( w_{\tau }-ikw\psi _{\tau }\right) , \\
  u_{\tau \tau } &=&\exp (-ik\psi )\left( w_{\tau \tau }-ikw\psi _{\tau \tau}-2ikw_{\tau }\psi _{\tau }-k^{2}w\psi _{\tau }^{2}\right).
\end{eqnarray*}
Explicitly, the evolution equation for $w$ is given by
\begin{eqnarray}\label{eq:10}
  w_{\xi} &=& -iaw_{\tau \tau }-ih\left\vert w\right\vert ^{2}w + ik\eps^2\frac{c-3e-2ak}{2} \left\vert w\right\vert ^{4}w \nonumber \\ 
  && - (c+2ak)\eps \left\vert w\right\vert w_{\tau }-(e+2ak)\eps w^{2}w_{\tau }^{\ast }-if\eps w\Hilb(\left\vert w\right\vert ^{2})_{\tau }.
\end{eqnarray}
If the free parameter $k^{\ast}$ is chosen as
\[
  k^{\ast} = -\frac{e}{2a}\eps,
\] 
equation (\ref{eq:10}) simplifies to
\begin{equation}\label{eq:11}
 w_{\xi }=-iaw_{\tau \tau }-ih\left\vert w\right\vert ^{2}w-i\frac{ce-2e^{2}}{4a}\eps^{2} \left\vert w\right\vert ^{4}w-(c-e)\eps \left\vert
 w\right\vert ^{2}w_{\tau }-if\eps w\Hilb(\left\vert w\right\vert^{2})_{\tau},
\end{equation}
which admits the Hamiltonian structure
\begin{equation}
 \left( 
 \begin{array}{c}
  w_{\xi } \\ 
  w_{\xi }^{\ast }
 \end{array}
 \right) = i\left( 
 \begin{array}{cc}
  0 & 1 \\ 
  -1 & 0%
 \end{array}
 \right) \left( 
  \begin{array}{c}
   \oD{\H_{w}}{w} \\ 
   \oD{\H_{w}}{w^{\ast}}
 \end{array}
 \right) ,
\end{equation}
where the Hamiltonian is given by
\begin{eqnarray}\label{eq:12}
  \H_{w} &=& \int \Bigl( a\left\vert w_{\tau }\right\vert ^{2}-\frac{h}{2}
 \left\vert w\right\vert ^{4}-\frac{ce-2e^{2}}{12a}\eps ^{2}\left\vert
 w\right\vert ^{6} \nonumber \\
 && -i\frac{c-e}{4}\eps \left\vert w\right\vert
 ^{2}(w_{\tau }^{\ast }w - w_{\tau }w^{\ast }) - \frac{f}{2}\eps \left\vert
 w\right\vert ^{2}\Hilb(\left\vert w\right\vert ^{2})_{\tau }\Bigr) d\tau.
\end{eqnarray}
Here, $\H_{w}$ is also an invariant together with the momentum 
\begin{equation}\label{eq:M}
  \M_{w} = \int i(w_{\tau }^{\ast }w-w_{\tau }w^{\ast })d\tau .
\end{equation}%
Note that in the Appendix of \cite{Trulsen1997} it was shown that Eq. \ref{eq:envB} admits the same invariant (\ref{eq:M}) if $\beta = 0$. Since $\H_{w}$ is an invariant for $w$, from (\ref{eq:9}) and (\ref{eq:M}) 
\begin{eqnarray}\label{eq:13}
 \E(u) &=&\H_{w}(\G^{-1}(w)) =
 \int \Bigl( a\left\vert u_{\tau }\right\vert ^{2}-\frac{h}{2}\left\vert
 u\right\vert ^{4}+\frac{ce+e^{2}}{6a}\eps ^{2}\left\vert u\right\vert
 ^{6} \\
 && -i\frac{c+e}{4}\eps \left\vert u\right\vert ^{2}(u_{\tau }^{\ast
 }u-u_{\tau }u^{\ast })-\frac{f}{2}\eps \left\vert u\right\vert^{2}\Hilb(\left\vert u\right\vert ^{2})_{\tau }\Bigr) d\tau ,  \nonumber \\
 &&  \nonumber \\
 \M(u) &=&\M_{w}(\G^{-1}(w))= \int \left[ i(u_{\tau }^{\ast }u-u_{\tau }u^{\ast })-\frac{e}{a}\eps\left\vert u\right\vert ^{4}\right] d\tau ,  \nonumber
 \end{eqnarray}%
are both invariants of the original spatial Dysthe equation (\ref{eq:we}). Unfortunately, $\E$ is not the associated Hamiltonian. In section \ref{sec:twaves} we present a highly accurate Fourier-type pseudo-spectral method that has been used to solve for the envelope dynamics and validate the invariance of the Hamiltonian (\ref{eq:12}) of $w$ and the new invariants (\ref{eq:13}) of $u$. In our numerical investigations, it is found that they are all conserved up to machine precision.

Note the appearance of terms of $O(\eps^{2})$ in the invariants $\E$ and $\H_{w}$. They both vanish if $e$ is null; as a result $u=w$ and $\E(u) = \H_{w}$ becomes the Hamiltonian for $u$. As a consequence, the spatial Dysthe (\ref{eq:SK}) for the wave envelope $B$ is non Hamiltonian since $e=2$, whereas that for the wave potential $A$ is (cf. Eq. \ref{eq:PO})).

To the leading order, the free-surface $\eta$ is related to the wave envelope $A=\eps u$ as
\[ k_{0}\eta \sim \eps ue^{i(k_{0}x-\omega _{0}t)}, \]
where from (\ref{eq:9}), $u$ can be expressed in terms of the canonical $w$ as
\begin{equation}\label{eq:eta}
 k_{0}\eta \sim \eps we^{i(k_{0}x-\omega _{0}t - i\eps k^{\ast}|\psi|)}.
\end{equation}
Thus, via the gauge transformation (\ref{eq:9})\ the wavenumber shift due to self-resonant interactions naturally arises, and the envelopes $|u|$ and $|w|$ are identical. This may suggest (\ref{eq:eta}) as a canonical ansatz for multiple scale perturbations of the Euler equations. Further, $w$ satisfies the canonical equation (\ref{eq:11}), viz.
\begin{equation}\label{eq:xx}
  w_{\xi} = -iw_{\tau\tau} - i\left\vert w\right\vert ^{2}w 
  - 2i\eps^2|w|^4w - 6\eps|w|^2w_{\tau} - 2i\eps w\Hilb(|w|^2)_{\tau},
\end{equation}
with the Hamiltonian given by 
\begin{eqnarray}
 \H_{w} = \int\Bigl(|w_{\tau}|^{2}-\frac{1}{2}|w|^{4} 
 - i\frac{3}{2}\eps|w|^{2}(w_{\tau}^{\ast}w - w_{\tau}w^{\ast}) \nonumber \\
 - \eps|w|^{2}\Hilb(|w|^{2})_{\tau} - \frac{2}{3}\eps^{2}|w|^{6}\Bigr)d\tau, \label{eq:yy}
\end{eqnarray}
which is invariant together with the action (\ref{eq:Ac})\ and momentum (\ref{eq:M}). Note that the corresponding $O(\eps^2)$ terms in (\ref{eq:xx}) and (\ref{eq:yy}) can be legitimately neglected and the Hamiltonian properties are preserved \cite{Salmon1983, Shepherd1990}.

The noncanonical envelope $u$ satisfies, besides the action (\ref{eq:Ac}), the two new invariants 
\[
 \E(u) = \int \left( \left\vert u_{\tau }\right\vert ^{2}-\frac{1}{2}
 \left\vert u\right\vert ^{4}-i\frac{5}{2}\eps \left\vert u\right\vert
 ^{2}(u_{\tau }^{\ast }u-u_{\tau }u^{\ast })+\frac{10}{3}\eps
 ^{2}\left\vert u\right\vert ^{6}-\eps \left\vert u\right\vert
 ^{2}H(\left\vert u\right\vert ^{2})_{\tau }\right) d\tau,
\]
\[
 \M(u) = \int \left[ i(u_{\tau }^{\ast }u-u_{\tau }u^{\ast
 })-2\eps \left\vert u\right\vert^{4}\right] d\tau,
\]
Practically, terms of $O(\eps^2)$ can be neglected and the invariants are approximately satisfied.

The spatial Dysthe (\ref{eq:PO}) of the wave potential $A=\eps v$ is already Hamiltonian with
\[
  \H_{v} = \int\left(|v_{\tau}|^{2} - \frac{1}{2}|v|^{4} - 2i\eps|v|^{2}(v_{\tau}^{\ast}v - v_{\tau}v^{\ast}) - \eps|v|^{2}\Hilb(|v|^{2})_{\tau}\right)d\tau,
\]
since the term $e=0$. Besides $\H_{v}$, it also conserves the action (\ref{eq:Ac}) and momentum (\ref{eq:M}) written in terms of $v$.

\section{Ground states and Traveling waves}

Insights into the underlying dynamics of the Dysthe equations are to be gained if we construct some special families of solutions in the form of ground states and traveling waves, often just called solitons or solitary waves. Hereafter, we do so for the equation (\ref{eq:we}) for the envelope $u$, and the associated Hamiltonian form (\ref{eq:11}) in $w$ can be treated in a similar way. However, owing to the gauge transformation (\ref{eq:9}) the envelope $|w| = |u|$.

First, consider the special case of (\ref{eq:we}) with parameters $(1,h,0,0,0)$. This is the classical NLS equation, which is well known to admit a family of localized traveling waves of the form (\cite{Zakharov1972}; see also \cite{Akhmediev1997, Sulem1999})
\begin{equation}\label{eq:NLSgs}
  u(\xi, \tau) = A\sqrt{\frac{2}{h}}\frac{e^{is\tau/2}}%
  {\cosh\bigl(A(\tau - s\xi)\bigr)}e^{-i\mu\xi}, \quad
  \mu = A^2 - \frac{s^2}{4},
\end{equation}
where $A$ is a free parameter. This solitary wave travels along the space $\xi$ with speed $1/s$ while pulsating in time. Due to the integrability of the NLS equation the interaction of solitary waves is elastic and so after the collision their shape is unchanged, but a phase shift occurs (see, for example, Figure \ref{fig:fig5}). At infinite speed, i.e. $s = 0$, (\ref{eq:NLSgs}) yields the so-called ground state solutions (see, for example, \cite{Yang2010}).

In the following, we wish to find traveling waves of the Dysthe equation (\ref{eq:we}) in $u$. To do so, we consider a generalization of (\ref{eq:NLSgs}) via the ansatz
\begin{equation}\label{eq:twsol}
  u(\xi, \tau) = v(\tau - s\xi)e^{i\mu\xi},
\end{equation}
where $\mu$ and $s$ are generic parameters and the function $v(\cdot)$ is in general complex. Substituting (\ref{eq:twsol}) into (\ref{eq:we}) and separating the real and imaginary parts $v = \Phi + i\Psi$ yield
\begin{eqnarray}\label{eq:Petv1}
  \mu\Phi - a\Phi_{\tau\tau} + s\Psi_\tau &=& h\Phi(\Phi^2+\Psi^2) + c\Psi_\tau(\Phi^2+\Psi^2) \nonumber \\ 
  && + e\Bigl(2\Phi\Phi_\tau\Psi - \Phi^2\Psi_\tau + \Psi^2\Psi_\tau\Bigr) + f\Phi\Hilb\bigl((\Phi^2+\Psi^2)_\tau\bigr), \\
  \mu\Psi - a\Psi_{\tau\tau} - s\Phi_\tau &=& h\Psi(\Phi^2+\Psi^2) - c\Phi_\tau(\Phi^2+\Psi^2) \nonumber \\
  && - e\Bigl(\Phi^2\Phi_\tau - \Phi_\tau\Psi^2 + 2\Phi\Psi\Psi_\tau\Bigr) + f\Psi\Hilb\bigl((\Phi^2+\Psi^2)_\tau\bigr). \label{eq:Petv2}
\end{eqnarray}
These equations can be solved analytically if the $c$, $e$ and $f$ terms can be assumed as small perturbations of the NLS equation. In this particular case direct soliton perturbation theory can be applied (see, for example, \cite{Akylas1989} and \cite{Yang2010}). We prefer to deal with the general case, and thus numerically solve for (\ref{eq:Petv1}) and (\ref{eq:Petv2}) using the Petviashvili method (\cite{Petviashvili1976}, see also \cite{Lakoba2007, Yang2010}). This numerical approach has been successfully applied by \cite{Zakharov2010} to compute ground states of their version of the temporal Dysthe equation.

To shorten the notation we rewrite the differential equations (\ref{eq:Petv1}) and (\ref{eq:Petv2}) in the operator form separating linear and nonlinear terms:
\[
  \L\left(
    \begin{array}{c}
      \Phi \\
      \Psi
    \end{array}\right) = \N(\Phi,\Psi),
\]
where $\N(\Phi,\Psi)$ denotes the right-hand side of equations (\ref{eq:Petv1}), (\ref{eq:Petv2}) and the matrix $\L$ is defined as
\[
  \L = \left(
    \begin{array}{cc}
      \mu - a\partial_{\tau\tau} & s\dtau \\
      -s\dtau & \mu - a\partial_{\tau\tau}
    \end{array}
  \right), \qquad
  \hat{\L} = 
  \left(
  \begin{array}{cc}
      \mu + ak^2 & isk \\
      -isk & \mu + ak^2
  \end{array}
  \right),
\]
where $\hat{\L}$ is the symbol of the operator $\L$ and $k$ is the Fourier transform parameter. Then the Petviashvili iteration takes the following form:
\[
  \left(
  \begin{array}{c}
      \Phi_{n+1} \\
      \Psi_{n+1}
  \end{array}
  \right) = \L^{-1}\cdot\N(\Phi_n,\Psi_n)\cdot
  \left(\frac{\left\langle\left(\begin{array}{c}
      \Phi_n \\
      \Psi_n
  \end{array}\right),\; \L\cdot\left(\begin{array}{c}
      \Phi_n \\
      \Psi_n
  \end{array}\right)\right\rangle}%
  {\left\langle\left(\begin{array}{c}
      \Phi_n \\
      \Psi_n
  \end{array}\right),\; \N(\Phi_n,\Psi_n)\right\rangle}\right)^{\gamma},
\]
where the exponent $\gamma$ is usually defined as a function of the degree of nonlinearity $p$ ($p = 3$ for the Dysthe equation (\ref{eq:we}), and $p = 5$ for (\ref{eq:11})). The rule of thumb prescribes the following formula $\gamma = \frac{p}{p-1}$. The scalar product is defined in the $L_2$ space. The inverse operator $\L^{-1}$ in the Fourier space is given by the following matrix:
\[
  \hat{\L}^{-1} = \frac{1}{(\mu + ak^2)^2 - (sk)^2}
  \left(
  \begin{array}{cc}
    \mu + ak^2 & -isk \\
      isk & \mu + ak^2
  \end{array}
  \right).
\]
To initialize the iterative process, one can use the analytical solution to the NLS (\ref{eq:NLSgs}) for the real part $\Phi_0$ and taking the imaginary part $\Psi_0 \equiv 0$. This iterative procedure is continued until the $L_\infty$ norm between two successive iteration is of the order of the machine precision. We would like to underline the fact that this method can be very efficiently implemented using the Fast Fourier Transform (FFT) (see, for example, \cite{Frigo2005}).

\begin{figure}
  \centering
  \includegraphics[width=0.83\textwidth]{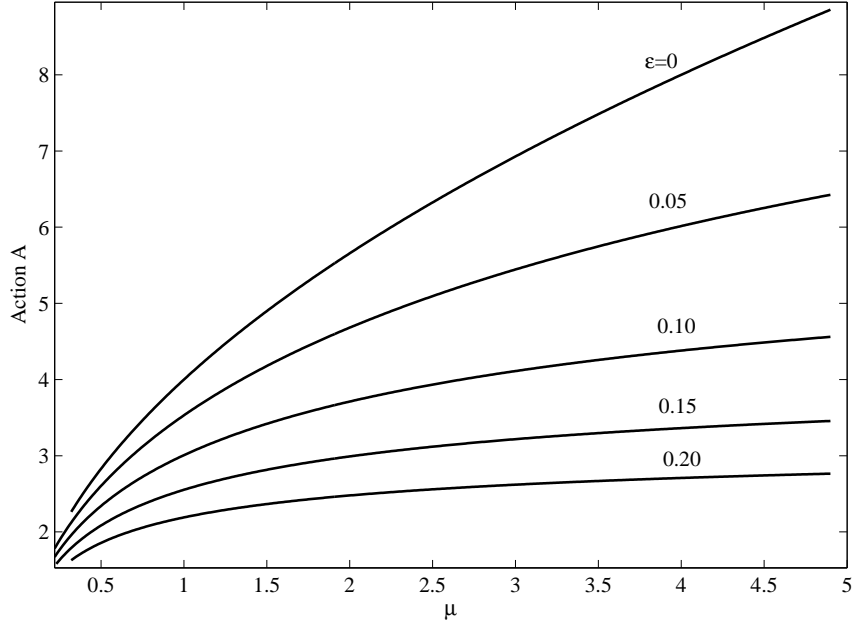}
  \caption{Action $\A$ of the ground state of the spatial Dysthe equation (\ref{eq:SK}) for $u$ as function of $\mu$, for different values of the steepness $\eps$.}
  \label{fig:fig1}
\end{figure}

\begin{figure}
  \centering
  \includegraphics[width=0.99\textwidth]{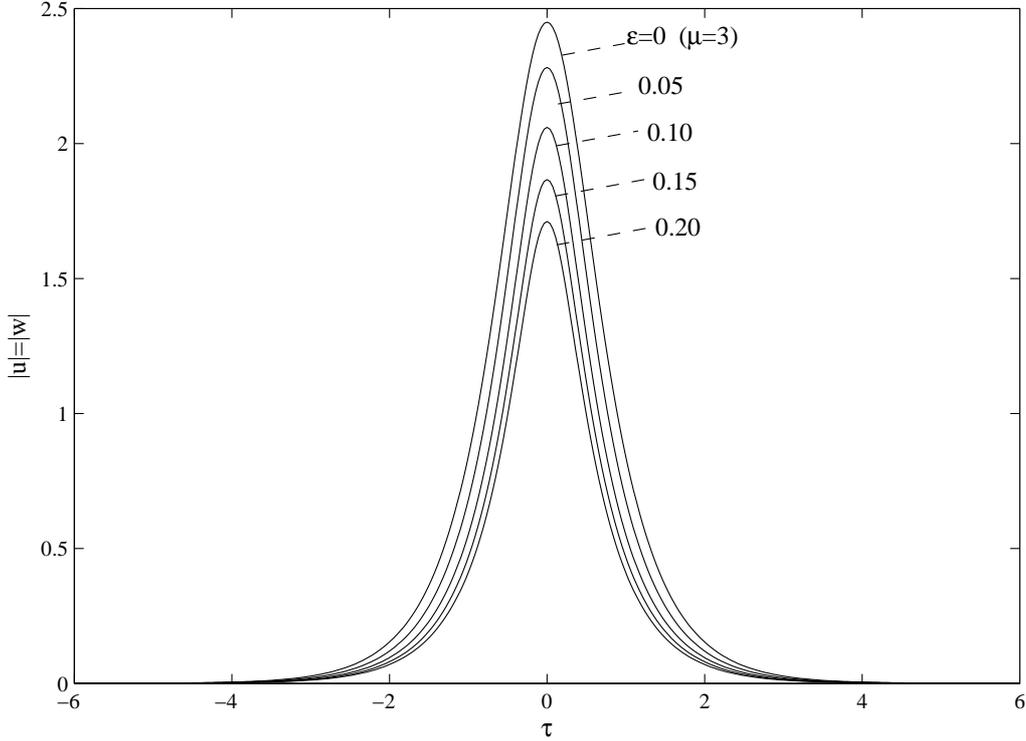}
  \caption{Envelopes $|u|$ of ground states of the spatial Dysthe equation (\ref{eq:SK}) for different values of the steepness $\eps$ ($\mu = 3$). The associated canonical envelope $|w| = |u|$ because of the gauge transformation (\ref{eq:9}).}
  \label{fig:fig2}
\end{figure}

As an application, consider the non-Hamiltonian Dysthe (\ref{eq:SK}), particular case of (\ref{eq:we}) with parameters $(1,1,8,2,2)$. Figure \ref{fig:fig1} shows the action $\A$ of the ground state ($s=0$) for $u$ (and so $w$ due to the gauge transformation (\ref{eq:9})) as function of $\mu$, computed for different values of the steepness $\eps$, and Figure \ref{fig:fig2} reports the associated envelopes $|u| = |w|$ for $\mu = 3$. As one can see, as $\eps$ increases they tend to reduce in size in agreement with the asymptotic analysis carried out by \cite{Akylas1989} in the limit of $\eps\to 0$.

\begin{figure}
  \centering
  \includegraphics[width=0.93\textwidth]{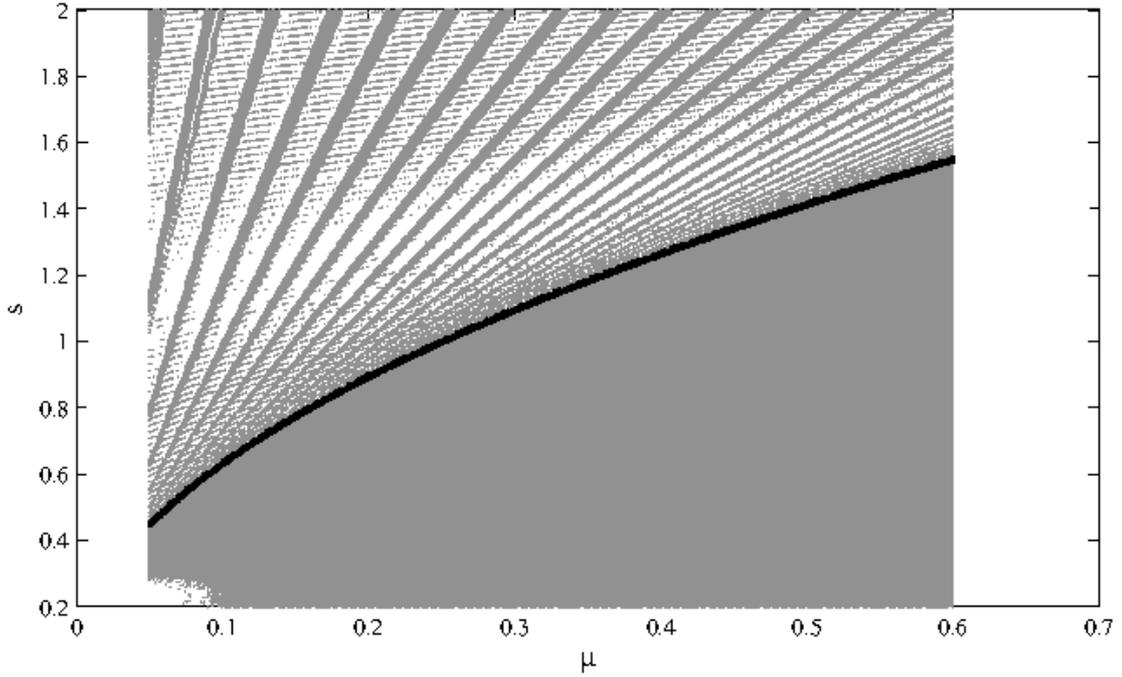}
  \caption{Numerical basin of attraction of the Petviashvili scheme in the phase space $(s,  \mu)$ for $\eps = 0.15$. Each gray dot corresponds to a well converged solution up to machine precision, whereas white spots are associated to divergent or converged-to-zero solutions. Solitary waves (localized traveling waves) occur below the black boundary curve $\Gamma$, which separates the region of periodic waves.}
  \label{fig:fig3}
\end{figure}

\begin{figure}
  \centering
  \includegraphics[width=0.99\textwidth]{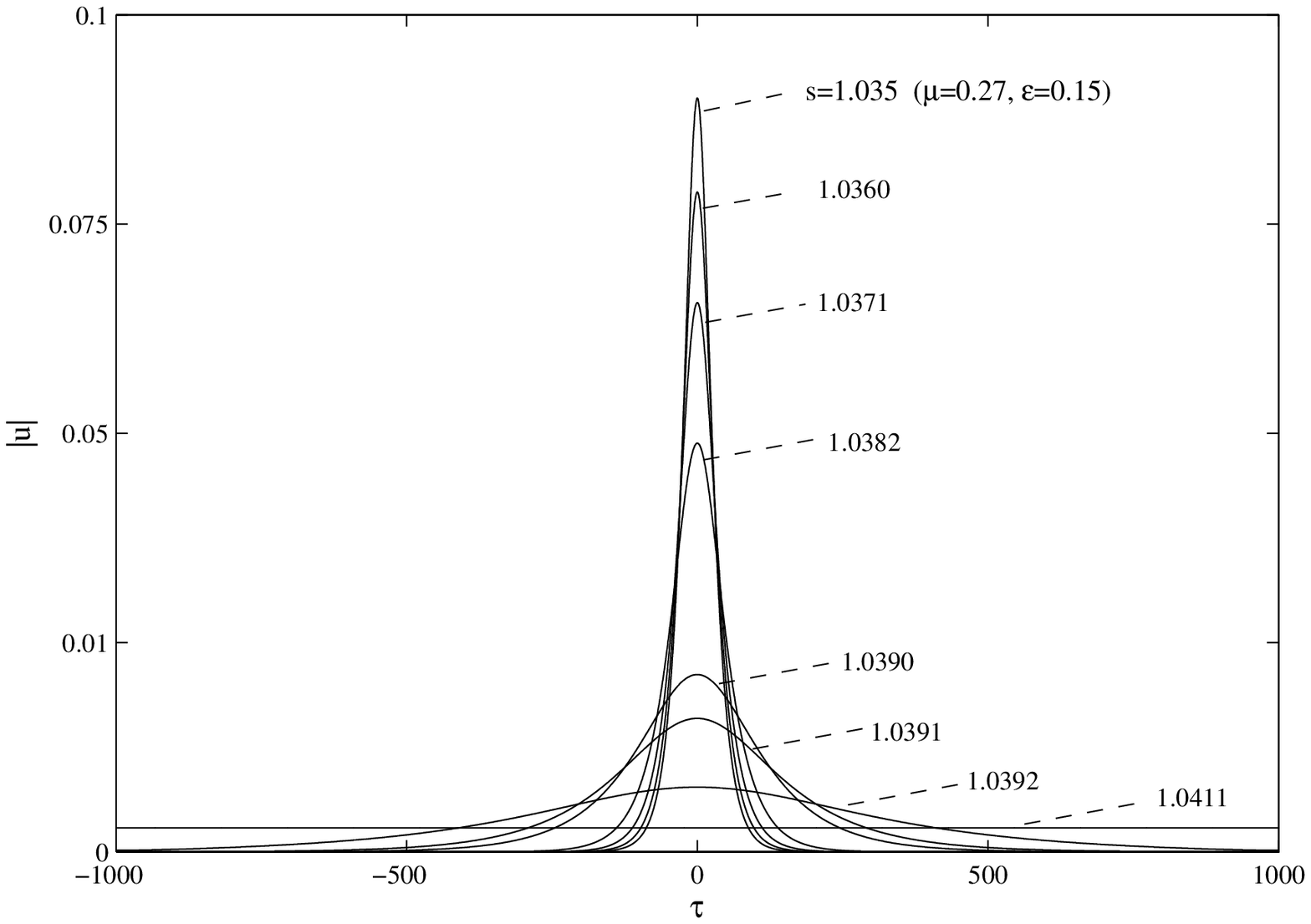}
  \caption{Localized traveling waves bifurcating from the boundary $\Gamma$ of Figure \ref{fig:fig3}. Change in shape of the envelope $|u|$ as $s$ varies while keeping $\mu = 0.27$. In particular, with reference to Figure \ref{fig:fig3}, the point $(s, \mu = 0.27)$ reaches the boundary $\Gamma$ from below.}
  \label{fig:fig4}
\end{figure}

Furthermore, Figure \ref{fig:fig3} illustrates a typical basin of attraction of the Petviashvili scheme in the phase space $(s, \mu)$ for $\eps = 0.15$. Each gray dot corresponds to a well converged solution up to machine precision, whereas white spots are associated to either divergent or converged-to-zero solutions. In particular, we noted that the numerical scheme converged to localized traveling waves below the black boundary curve $\Gamma$ shown in Figure \ref{fig:fig3}. On the other hand, the convergence to simple periodic waves occurred for points above $\Gamma$. An analytical form of such curve can be derived from the nonlinear dispersion relation $A$ -- $\omega$ of periodic waves of the form
\[
  u(\xi,\tau) = Ae^{i\omega(\tau-s\xi)}e^{-i\mu\xi}.
\]
Indeed, using this ansatz in (\ref{eq:we}) yields
\begin{equation}\label{eq:quadr}
  a\omega^{2} + (s+(e-c)A^2)\omega + \mu - hA^{2} = 0.
\end{equation}
The boundary $\Gamma$ of the periodic waves region in the phase space $(s, \mu)$ follows from the condition that the discriminant of the quadratic equation in $\omega$ is null, viz.
\begin{equation}\label{eq:sep}
  s = (c-e)A^{2} + 2\sqrt{a(\mu - hA^2)}.
\end{equation}
The analytical separatix (\ref{eq:sep}) is practically identical to the boundary $\Gamma$ that we have identified numerically, confirming the validity and correctness of our numerical solver. Thus, localized traveling waves bifurcate from $\Gamma$. This is clearly illustrated in Figure \ref{fig:fig4}, which reports the change in shape of the envelope $|u|$ as $s$ varies while keeping $\mu$ constant, say $\mu = 0.27$ as a particular case. Clearly, as the point $(s, \mu = 0.27)$ reaches the boundary $\Gamma$ from below (see Figure \ref{fig:fig3}), the soliton envelope tends to flatten to that of a periodic wave. Similar results also hold for the Hamiltonian $w$ due to the gauge transformation (\ref{eq:9}).

\section{Traveling wave interactions}\label{sec:twaves}

We further investigate the dynamics of the Dysthe equations by means of a highly accurate Fourier-type pseudo-spectral method, which is needed in order validate the newly derived conservation laws and investigate nonlinear phenomena such as elastic/inelastic traveling waves interactions. Spectral methods have been already proven to be sufficiently accurate and robust for studying the dynamics of gravity waves \cite{Milewski1999, Clamond2001, Yang2010}. As in the previous section, we first illustrate the numerical method for equation (\ref{eq:we}), but the associated Hamiltonian form (\ref{eq:11}) can be treated in a similar way. Then we present some results of the interactions among traveling waves.

\subsection{Numerical method description}

Equation (\ref{eq:we}) can be written in the following operator form:
\begin{equation}\label{eq:Dys}
  u_\xi + \L\cdot u = \N(u),
\end{equation}
where the operators $\L$ and $\N$ are defined as
\[
  \L = ia\partial_{\tau\tau}, \qquad
  \N(u) = - ih|u|^{2}u - c\eps|u|^2u_{\tau} - \eps e u^2u_{\tau}^{\ast} 
  - if\eps u \Hilb(|u|^2_\tau).
\]
Equation (\ref{eq:Dys}) will be solved numerically by applying the Fourier transform in the time variable $\tau$. The transformed variable will be denoted by $\hat{u} = \F\{u\}$. We recall that the symbol of the last term $\Hilb(\partial_{\tau})$ is equal to $|k|$, and that of $\L$ is $-iak^2$, $k$ being the Fourier transform parameter. The nonlinear terms are computed in the physical space, while time derivatives and the Hilbert transform are computed in the Fourier space. For example, the term $|u|^2u_\tau$ is discretized as:
\[
  \F\{|u|^2u_\tau\} = \F\{|\F^{-1}(\hat{u})|^2\cdot\F^{-1}\{ik\hat{u}\}\}.
\]
The other nonlinear terms are treated in a similar way. We note that the usual 4-half rule is applied for anti-aliasing \cite{Trefethen2000, Clamond2001, Fructus2005}.

In order to improve the stability of the space discretization procedure, we can integrate exactly the linear terms. This is achieved by making a change of variables \cite{Milewski1999, Fructus2005}:
\[
  \hat{v}_\xi\ =\ e^{(\xi-\xi_0)\L}\cdot\N\Bigl\{e^{-(\xi-\xi_0)\L}\cdot\hat{v}\Bigr\},
   \qquad 
  \hat{v}(\xi)\ \equiv\ e^{(\xi-\xi_0)\L}\cdot\hat{u}(\xi), \qquad 
  \hat{v}(\xi_0)\ =\ \hat{u}(\xi_0).
\]
Finally, the resulting system of ODEs is discretized in space by the Verner's embedded adaptive 9(8) Runge--Kutta scheme \cite{Verner1978}. The step size is chosen adaptively using the so-called H211b digital filter \cite{Soderlind2003, Soderlind2006} to meet the prescribed error tolerance, set as of the order of machine precision.

\subsection{Numerical results}

\begin{figure}
  \centering
  \includegraphics[width=1.17\textwidth]{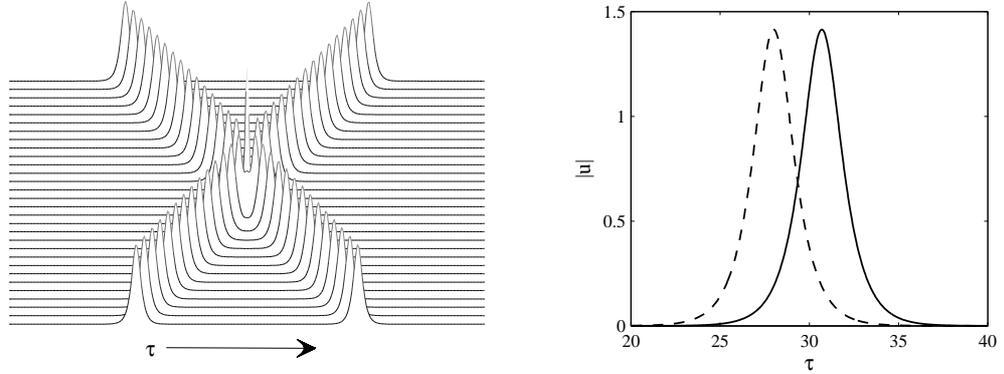}
  \caption{(Left) elastic collision of two NLS solitary waves traveling at the same speed $s=2$, for $\mu = 2$, and (right) initial ($--$) and final (---) shapes of one of the two solitons.}
  \label{fig:fig5}
\end{figure}

\begin{figure}
  \centering
  \includegraphics[width=1.17\textwidth]{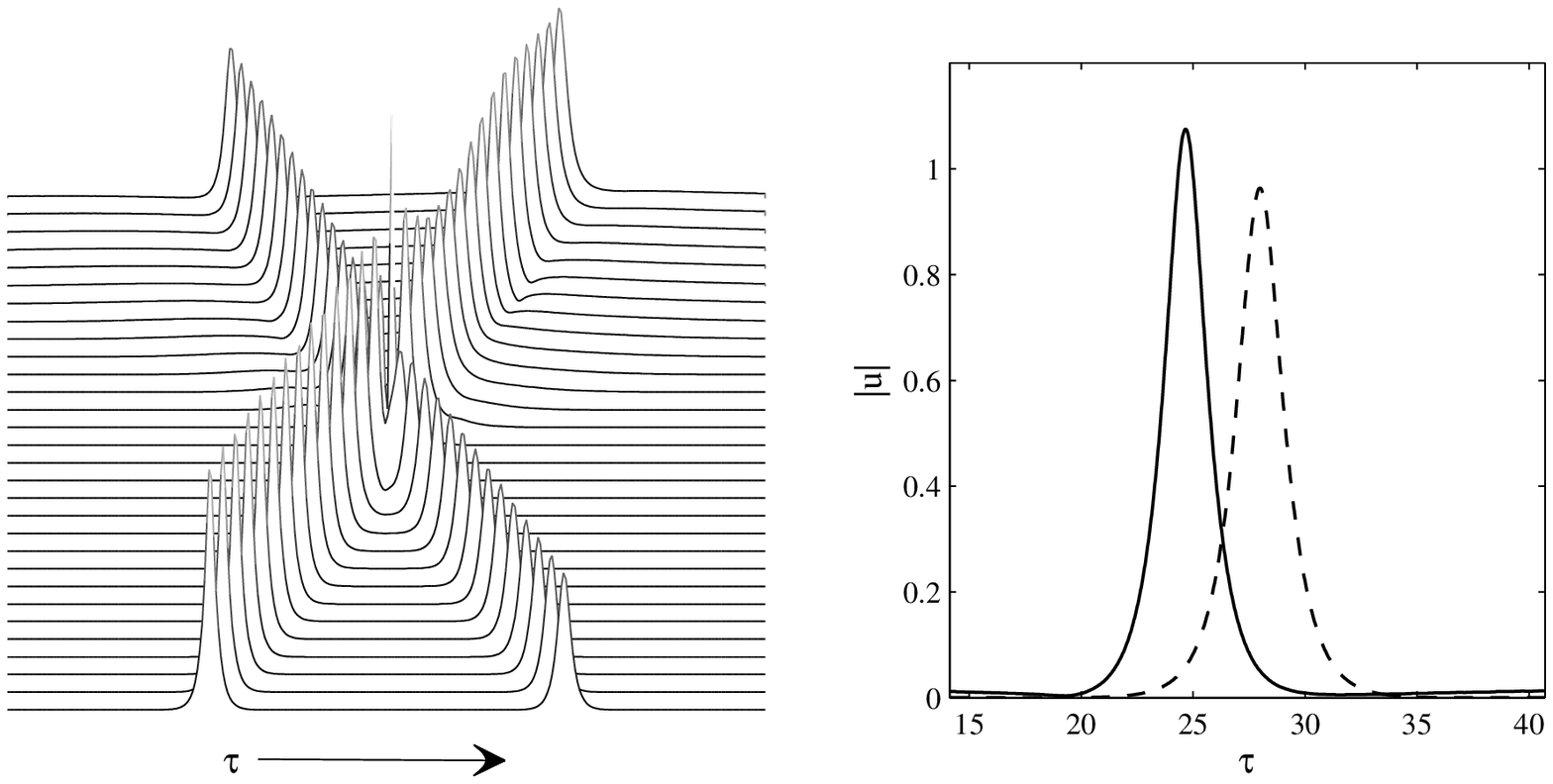}
  \caption{(Left) inelastic collision of two Dysthe solitary waves traveling at the same speed $s=2$, for $\mu = 2$, and (right) initial ($--$) and final (---) shapes of one of the two solitons.}
  \label{fig:fig6}
\end{figure}

\begin{figure}
  \centering
  \includegraphics[width=0.99\textwidth]{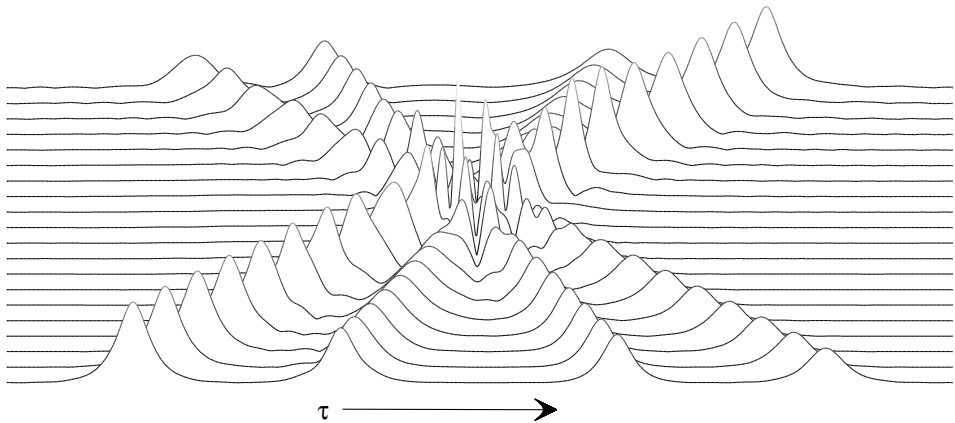}
  \caption{Inelastic collision of four Dysthe solitary waves.}
  \label{fig:fig7}
\end{figure}

As an application, consider the interaction between two solitary waves of the NLS equation traveling in opposing directions with the same speed $s = 2$, for $\mu = 2$ ($\eps = 0$). The shape of the two solitons is identical because of the NLS reflection symmetry, i.e. $u(\xi, \tau) = (\xi, -\tau)$. The left plot of Figure \ref{fig:fig5} shows that the two solitons emerge out of the collision with the same shape, but a phase shift. The interaction is elastic as it should be since the NLS equation is integrable. This is clearly seen from the right plot of the same figure, which reports the initial and final shapes of one of the two solitary waves. We point out that the Hamiltonian and the other two NLS invariants are conserved up to machine precision. For the same parameters, Figure \ref{fig:fig6} reports the interaction of the associated solitary waves of the Dysthe equation (\ref{eq:SK}) for $\eps = 0.15$. The reflection symmetry is lost and the two solitons have different shape and amplitude. The interaction is clearly inelastic, since after the collision radiation is shed and the initial and final soliton shapes are different as seen in the right plot of Figure \ref{fig:fig6}. The interaction of four solitons has similar inelastic characteristics as shown in Figure \ref{fig:fig7}. This suggests the non-integrability of the Dysthe equation (\ref{eq:SK}). Similar dynamics is also observed for the associated Hamiltonian form (\ref{eq:11}).

\section{Conclusions}

A canonical variable for the spatial Dysthe equation for the wave envelope $u$ has been identified by means of the gauge transformation (\ref{eq:9}), and the hidden Hamiltonian structure is unveiled. Further, the associated equation for the wave potential $v$ is shown to be already Hamiltonian. Exploiting the gauge invariance allowed to derive two new invariants for the noncanonical $u$. Clearly (\ref{eq:9}) is also applicable to the temporal Dysthe equation (\ref{eq:envB}), but it leads to a cumbersome equation in $w$ that contains nonlinear terms with derivatives  higher than the first order, which cannot be eliminated by a proper choice of the free parameter $k$. The transformed equation may well be Hamiltonian, but we could not identify such functional due to the complexity of the equation. In any case, the gauge transformation (\ref{eq:9}) does not yield the Hamiltonian form derived by \cite{Gramstad2011} for the temporal case.

Moreover, the existence of localized traveling waves that bifurcate from periodic waves has been investigated numerically by means of a highly accurate Petviashvili scheme. In particular, ground state solutions are found to be in agreement with the asymptotic analysis of \cite{Akylas1989}. A Fourier-type pseudo-spectral method has been developed to solve for the envelope dynamics and test the new conservation laws, which are validated up to machine precision. Solitary waves are found to interact inelastically, suggesting the non-integrability of both the Hamiltonian and non-Hamiltonian version of the spatial Dysthe equations.

\section*{Acknowledgements}

D.~Dutykh acknowledges the support from French Agence Nationale de la Recherche, project MathOc\'ean (Grant ANR-08-BLAN-0301-01). F.~Fedele acknowledges the travel support received by the Geophysical Fluid Dynamics (GFD) Program to attend part of the summer school on ``Shear Turbulence: Onset and Structure'' at the Woods Hole Oceanographic Institution in August 2011. 

We would like to thank also Professors Didier Clamond, Taras Lakoba, Paul Milewski, Lev Shemer and Jianke Yang for useful discussions on the subject of nonlinear waves.

F.~Fedele thanks Professor Phil Morrison for useful discussions and a short lecture on Hamiltonian systems during the 2011 GFD program.

\bibliography{biblio}
\bibliographystyle{plain}

\end{document}